\documentclass[a4paper,twocolumn,floatfix]{revtex4}
\usepackage{graphicx}

\begin{document}

\title{Accelerating, hyper-accelerating, and decelerating probabilistic networks}

\author{M. J. Gagen and J. S. Mattick}

\affiliation{ARC Special Research Centre for Functional and
Applied Genomics, Institute for Molecular Bioscience, University
of Queensland, Brisbane, Qld 4072, Australia}

\email{m.gagen@imb.uq.edu.au}

\date{\today}

\begin{abstract}
Many growing networks possess accelerating statistics where the
number of links added with each new node is an increasing function
of network size so the total number of links increases faster than
linearly with network size. In particular, biological networks can
display a quadratic growth in regulator number with genome size
even while remaining sparsely connected.  These features are
mutually incompatible in standard treatments of network theory
which typically require that every new network node possesses at
least one connection. To model sparsely connected networks, we
generalize existing approaches and add each new node with a
probabilistic number of links to generate either accelerating,
hyper-accelerating, or even decelerating network statistics in
different regimes. Under preferential attachment for example,
slowly accelerating networks display stationary scale-free
statistics relatively independent of network size while more
rapidly accelerating networks display a transition from scale-free
to exponential statistics with network growth.  Such transitions
explain, for instance, the evolutionary record of single-celled
organisms which display strict size and complexity limits.
\end{abstract}

\maketitle

\section{Introduction}

The rapidly expanding field of network analysis, reviewed in
\cite{Dorogovtsev_02_10,Albert_02_47,Newman_03_167}, has provided
examples of networks exhibiting ``accelerating" network growth,
where total link number grows faster than linearly with network
size \cite{Dorogovtsev_01_056125}. For instance, the Internet and
World Wide Web appear to grow by adding links more quickly than
sites \cite{Faloutsos_99_25,Dorogovtsev_01_025101} though the
relative change over time is small \cite{Vazquez_02_066130}.
Similarly, both the number of links per substrate in the metabolic
networks of organisms \cite{Jeong_00_65} and the average number of
links per scientist in collaboration networks increases linearly
over time
\cite{Dorogovtsev_00_33,Vazquez_00_0006132,Barabasi_01_0104162,Barabasi_02_590,Vazquez_03_056104},
while also languages appear to evolve via accelerated growth
\cite{Dorogovtsev_01_2603}. These studies have motivated
examinations of stochastic accelerating networks where the number
of new links added is probabilistic (though integral and greater
than one) causing network transitions from scale-free to
exponential statistics
\cite{Albert_00_5234,Krapivsky_01_5401,Liu_02_036112,Goh_02_108701,Vazquez_03_056104,Sen_0310513}.

These previous studies of deterministic or stochastic networks
have typically considered networks where every node has at least
one connection which constrains the rate of acceleration that can
be considered. In particular, if each new node added to a network
of $N$ nodes is accompanied by $N^\alpha\geq 1$ new links with
acceleration parameter $\alpha\geq 0$, then ensuring that the
network is less than fully connected constrains acceleration
parameters to the range $0\leq\alpha<1$
\cite{Dorogovtsev_01_025101}. Equivalent limits were considered in
Ref. \cite{Sen_0310513}. However, real networks can contain a
majority of nodes that are entirely unconnected---many computers
are unconnected or only intermittently connected to the Internet
and many people are unconnected nodes in social or sexual contacts
networks. Modelling networks where a substantial proportion of the
nodes are unconnected requires a probabilistic approach in which
the number of links added with each new node scales as $pN^\alpha$
for some probability parameter $0\leq p\leq 1$.  The introduction
of a probabilistic envelope lifts constraints on the acceleration
parameter allowing networks displaying deceleration $\alpha<0$, no
acceleration $\alpha=0$, acceleration $0\leq\alpha<1$, and
hyper-acceleration $\alpha\geq 1$.

Hyper-accelerating probabilistic networks with $p\ll 1$ and
$\alpha\geq 1$ are not merely of theoretical interest, and appear
for instance in prokaryote gene regulatory networks where many
gene nodes are unregulated and merely constitutively or
stochastically expressed \cite{Gagen_0312021,Gagen_0312022}.
Prokaryote gene regulatory networks are sparsely connected ($p\ll
1$) and display hyper-acceleration $\alpha=1$ as established by
independent comparative genomics analyses
\cite{Stover_00_959,Bentley_02_141,vanNimwegen_03_479,Cases_03_248,Croft_03_unpub}.
This is likely due to their reliance on sequence homology
interactions between protein transcription factors and specific
promoter binding sequences sites \cite{Croft_03_unpub}. In these
regulatory networks, outbound regulatory links are preferentially
attached to existing regulator nodes as gene duplication events
contribute to up to 75\% of each of new genes
\cite{Yanai_00_2641,Qian_01_673,Wagner_02_457,Eisenberg_03_138701,PastorSatorras_03_199,Giot_03_1727,Kunin_03_1589}
and regulatory transcription factors
\cite{MandanBabu_03_1234,Bhan_02_1486}.  In contrast, inbound
links to regulated nodes are randomly formed due to the random
drift of gene promotor sequences, although their subsequent
fixation is determined by selection.  The high acceleration
parameter $\alpha=1$ ensures that these prokaryote gene regulatory
are size and complexity constrained at predicted limits which
closely match genomic size limits observed in the evolutionary
record \cite{Croft_03_unpub,Gagen_0312021,Gagen_0312022}.
Accelerating networks are more prevalent and important in society
and in biology than is commonly realized. In fact, any network in
which the dynamical state of any node depends on the immediately
preceding state of (ideally) every other node is accelerating and
will perhaps display either structural transitions from randomly
connected, to scale free statistics, to densely connected and
perhaps finally to fully connected statistics, or explicit size
and complexity constraints \cite{Mattick_05_856}.

This paper generalizes previous analysis to consider probabilistic
accelerating networks in decelerating regimes with $\alpha<0$,
accelerating regimes with $0\leq\alpha<1$, and hyper-accelerating
regimes where $\alpha>1$.  We will show that hyper-accelerating
networks under preferential attachment of new connections to
existing nodes can display transitions in their connectivity
statistics from scale-free to exponential statistics dependent on
the acceleration parameter $\alpha$, the probability parameter
$p$, and network size $N$.  These transitions occur at the point
where the accelerating growth in connection number can no longer
be sustained introducing exponential cut-offs in the connectivity
distributions.  These cutoffs can be generated by different
mechanisms and appear in growing networks subject to strong aging
effects \cite{Dorogovtsev_00_1842}, in finite sized networks
\cite{AutoMoreira_02_268703}, and in information filtered networks
\cite{Mossa_02_138701,Stefancic_04_0404495}.

To complete our examination of network transitions in
accelerating, hyper-accelerating and decelerating probabilistic
networks, we first define our growing network models using
undirected links in Section \ref{sect_prob_accelerating_networks}.
This definition allows an immediate rough quantification of the
probability of an accelerating network forming a single giant
connected component or of undergoing a transition to a fully
connected regime with network growth. In Section
\ref{sect_pref_attachment} we examine probabilistic networks
growing through the preferential attachment of new links to
existing connected network nodes, while in Section
\ref{sect_random} we examine growing probabilistic networks
featuring the random attachment of new links to all existing
nodes.

\section{Probabilistic accelerating networks}
\label{sect_prob_accelerating_networks}

We consider networks growing through the sequential addition of
numbered nodes $n_k$ for $1\leq k\leq N$ where at network size
$k$, node $n_i$ ($1\leq i\leq k$) has $l_{ik}$ undirected links.
The addition of node $n_k$ and its $l_{kk}$ links will increase
the probable number of links attached to existing nodes $n_i$ for
$1\leq i\leq (k-1)$ so $l_{ik}\geq l_{ii}$. We assume new links
are added only with each arriving node (so no new links are added
between established nodes), and that the average number of new
links attached to node $n_k$ on its entry to the network is a
function of network size
\begin{equation}
   l_{kk} = p k^\alpha,
\end{equation}
where the probability constant $p$ satisfies $0\leq p\leq 1$. This
average is then a decreasing function of network size for
$\alpha<0$, constant for $\alpha=0$, and an increasing function of
network size for $\alpha>0$. The total expected number of links
$L$ in a probabilistic network of large size $N$ is then
\begin{eqnarray}    \label{eq_net_parameters}
    L &=& \sum_{k=1}^{N} pk^\alpha \;\approx\;
            \int_1^N pk^\alpha  \nonumber \\
    &\approx & \left\{
    \begin{array}{ll}
      \frac{-p}{1+\alpha} \left(1-N^{1+\alpha}\right) \;\approx\; -\frac{p}{1+\alpha}, & \alpha<-1, \\
       &  \\
      p \ln N, & \alpha=-1, \\
       &  \\
      \frac{p}{1+\alpha} \left(N^{1+\alpha}-1\right) \;\approx\; \frac{pN^{1+\alpha}}{1+\alpha} , & \alpha>-1. \\
    \end{array}
    \right.
\end{eqnarray}
Consequently, the expected number of connections is $C=2L$, and
the average connectivity per node is $\langle k\rangle=C/N$.  In
later numerical simulations, we compare the statistics of networks
of different sizes which is possible provided they possess the
same average connectivity.  We achieve this for accelerating
networks with $\alpha\geq 0$ by choosing different probability
parameters $p$ for each different sized network according to
\begin{equation}      \label{eq_bar_p}
  p = \frac{(1+\alpha)\langle k\rangle}{2N_c^\alpha}.
\end{equation}
This allows us to later consider different networks with identical
average connectivity $\langle k\rangle=0.4$, where networks have
either a critical size of $N_c=1000$ nodes connected by $L=200$
links with $p=0.2(1+\alpha)10^{-3\alpha}$, or a critical size of
$N_c=10,000$ nodes connected by $L=2,000$ links with
$p=0.2(1+\alpha)10^{-4\alpha}$.

Because link formation is probabilistic, the total number of links
attached to node $n_k$ on entry to the network, denoted $j$ say,
ranges between $0$ and $k$, so $0\leq j\leq k$ with average
$pk^{\alpha}$. As each link either forms or does not form, we
model the link formation process using a binomial distribution.
For the binomial distribution, the average $l_{kk}=pk^{\alpha}$
equates to the product of the maximum number of links $k$ and the
link formation probability. Conversely, the size dependent link
formation probability at network size $k$ equates to
$l_{kk}/k=pk^{\alpha-1}$. Consequently, the probability
distribution that node $n_k$ forms exactly $j$ links on entry to
the network is
\begin{equation}   \label{eq_P_j_k_dist}
 P(j,k) = {k \choose j} \left(pk^{\alpha-1}\right)^j
    \left[1-pk^{\alpha-1}\right]^{k-j}.
\end{equation}
That is, when $\alpha=0$ the latest node $n_k$ can potentially
form links with every one of the $k$ existing nodes (counting
itself) with a network size dependent probability $p/k$ to give an
average number of new links of $p$ connections with each new node.
When $\alpha=1$, the latest node $n_k$ can potentially form links
with every other node with constant probability $p$ to give an
average number of new links of $pk$ with each new node, and so on.
The connection probability is constrained to be less than unity,
$pk^{\alpha-1}\leq 1$, or equivalently, the average number of
links added to node $n_k$ is required to be less than $k$,
$pk^\alpha\leq k$. For constant $p$, hyper-accelerating networks
with $\alpha>1$ will eventually violate these constraints with
network growth which effectively imposes size and structural
constraints on these networks.

There are two types of network connectivity transitions which have
been of interest in the literature.  The first occurs when a
growing network changes from sparse to dense connectivity as this
change necessarily implies that the final connectivity statistics
will become exponential in nature. That is, while a sparsely
connected network can possess either scale-free or exponential
statistics, a densely connected network must exhibit exponential
statistics (as multiple connections are not allowed between any
two given nodes). This transition occurs for undirected links when
approximately $L>N^2/2$. The second transition of interest occurs
when the connected nodes coalesce to form a single giant island of
interconnected nodes. To borrow results from random graph theory
\cite{Erdos_60_17} (which have been roughly validated for
accelerating biological networks
\cite{Gagen_0312021,Gagen_0312022}), the giant island is expected
to form when approximately $L>N/2$.  We now investigate the
dependence of these transitions on the acceleration parameter
$\alpha$.

It is evident that decelerating networks with $\alpha<0$ are very
sparsely connected so neither of the dense connectivity or giant
island transitions will occur. Hereinafter, we only consider
accelerating networks with $\alpha\geq 0$. For such networks, the
dense connectivity and giant island transitions will occur
respectively at the points
\begin{eqnarray}
     L > \frac{N^2}{2} &\Rightarrow&
             N^{\alpha-1} > \frac{1+\alpha}{2p} =
             \frac{N_c^\alpha}{\langle k\rangle}, \nonumber \\
     L > \frac{N}{2} &\Rightarrow&
             N^{\alpha} > \frac{1+\alpha}{2p} =
             \frac{N_c^\alpha}{\langle k\rangle}.
\end{eqnarray}
This establishes that, for instance, non-accelerating networks
with $\alpha=0$ can never undergo a transition to a densely
connected regime for any choice of the probability parameter $p$
for large network sizes $N$, though these networks can form giant
islands with network growth as long as $p>1/2$.  This inability to
become densely connected remains for all accelerating regimes
$\alpha<1$, but dense connectivity can emerge in the
hyper-acceleration regime. At this point with $\alpha=1$, growing
networks can undergo a transition to the densely connected regime
when the probability parameter is large, $p\approx 1$.  For yet
higher acceleration rates $\alpha>1$, all growing networks undergo
a transition to the dense connectivity regime and display single
giant islands, and these transitions occur at network sizes
determined by the probability parameter $p$.

In the following sections, we determine the connectivity
statistics for accelerating and hyper-accelerating networks under
both preferential and random attachment of new links to
established nodes.

\section{Preferentially attached accelerating networks}
\label{sect_pref_attachment}

We now turn to consider accelerating and hyper-accelerating
networks ($\alpha\geq 0$) where new links are principally
preferentially attached to existing connected nodes.  The final
generated link distribution will depend on the balance of
different countervailing trends, namely the respective weights
given to preferential versus random attachment, and the magnitude
of the acceleration parameter which influences both the rate at
which links are attached to newer nodes and the rate at which
while older nodes accumulate links.  We now examine these trends
in detail.

On entry into the network, node $n_k$ establishes an average of
$l_{kk}=pk^\alpha$ links with existing nodes
$n_j\in\{n_1,\dots,n_k\}$. We suppose that these links either are
preferentially attached to node $n_j$ with probability
proportional to $n_j$'s current connectivity $l_{jk}$, or are
randomly attached to $n_j$ with probability proportional to
$\beta$, a random connection parameter. Using the continuous
approximation
\cite{Barabasi_99_17,Barabasi_99_50,Dorogovtsev_01_056125}, the
rate of growth in  link number for node $n_j$ is
\begin{equation}    \label{eq_continuum_links}
  \frac{\partial l_{jk}}{\partial k} =
     l_{kk} \frac{(l_{jk}+\beta)}{\int_{0}^{k} (l_{jk}+\beta) \; dj}.
\end{equation}
Here, the rate of growth in the connectivity $l_{jk}$ is
determined by the number of new links added with node $n_k$ (i.e.
$l_{kk}$) which can be either preferentially attached to the
existing nodes $n_j$ with probability proportional to that nodes
current connectivity (i.e. $l_{jk}$), or randomly attached if
$\beta>0$ allowing even initially unconnected nodes to establish
connections. (The case $\beta=0$ ensures that a node that receives
zero links on entry to the network remains unconnected for all
time.)

The denominator of Eq. \ref{eq_continuum_links} is a probability
weighting to ensure normalization and is roughly equal to the
total number of connections ($C$) for all nodes at network size
$k$. Following \cite{Dorogovtsev_02_10}, we can evaluate the
denominator using the identity
\begin{equation}
  \frac{\partial}{\partial k} \int_0^k l_{jk} \;dj =
     \int_0^k  \frac{\partial l_{jk}}{\partial k}  \; dj
     + l_{kk},
\end{equation}
which can be evaluated using Eq. \ref{eq_continuum_links} to give
\begin{equation}
  \frac{\partial}{\partial k} \int_0^k  l_{jk} \;dj =
  2 l_{kk}.
\end{equation}
Noting $l_{kk}\approx pk^\alpha$, this can be integrated to
determine the denominator of Eq. \ref{eq_continuum_links} to be
\begin{equation}
   \int_0^k ( l_{jk}+\beta) \;dj =
   \frac{2p}{\alpha+1}k^{\alpha+1}
   + \beta k.
\end{equation}
The first term on the RHS is the total number of connections at
network size $k$ and is in agreement with Eq.
\ref{eq_net_parameters}. Substituting this relation back into Eq.
\ref{eq_continuum_links} gives
\begin{equation}           \label{eq_partial_ljk}
  \frac{\partial l_{jk}}{\partial k} =
     \frac{p k^{\alpha-1} ( l_{jk}+\beta)}{\left(
     \frac{2p}{\alpha+1} k^\alpha + \beta \right)}.
\end{equation}
The connectivity statistics can readily be obtained by integrating
Eq. \ref{eq_partial_ljk} with the initial conditions
$l_{jj}\approx pj^\alpha$ at time $j$ to obtain
\begin{equation}     \label{eq_links_l_jn_beta}
  l_{jN} = \left\{
    \begin{array}{ll}
      (p+\beta) \left(\frac{N}{j}\right)^{\frac{p}{2p+\beta}}-\beta, & \alpha=0, \\
       &  \\
      (pj^\alpha+\beta)
       \left(\frac{\frac{2p}{\alpha+1}N^\alpha+\beta}{\frac{2p}{\alpha+1}j^\alpha+\beta}
       \right)^{\frac{\alpha+1}{2\alpha}} - \beta,   &  \alpha>0. \\
    \end{array} \right. \\
\end{equation}
Differentiation of these curves with respect to network size $N$
satisfies Eq. \ref{eq_continuum_links}, while integrating these
link numbers over all node numbers $j$ (numerically for
$\alpha>0$) gives the required total number of links as in Eq.
\ref{eq_net_parameters}. Further, differentiating the linkage
distribution $l_{jN}$ with respect to the age index $j$
establishes that average node connectivity is monotonically
decreasing with increasing node age $j$ for $0\leq \alpha<1$, flat
for $\alpha=1$, and monotonically increasing for $\alpha>1$. In
other words, irrespective of the choice of the $\beta$ parameter,
slowly accelerating networks with $0\leq \alpha<1$ have the
majority of their connections in the oldest portions of their
networks, while hyper-accelerating networks with $\alpha>1$ place
the majority of the connections in the youngest portions of their
networks. The transition point occurs at $\alpha=1$ when
connectivity is uniformly distributed over the network
irrespective of node age.

The further analysis of this age dependent connectivity
distribution is heavily dependent on the setting of the random
choice parameter $\beta$.  Self-evidently, setting $\beta$ small
ensures that preferential attachment processes dominate while the
choice of large $\beta$ ensures that random attachment dominates.
Appendix \ref{app_pref_random_attach} uses a Taylor series
expansion to demonstrate that preferential attachment dominates
hyper-accelerating networks only when $\beta$ is so small as to be
essentially zero, and hereinafter, we analyze preferential
attachment using the setting $\beta=0$.

\begin{figure}[htbp]
\centering
\includegraphics[width=\columnwidth,clip]{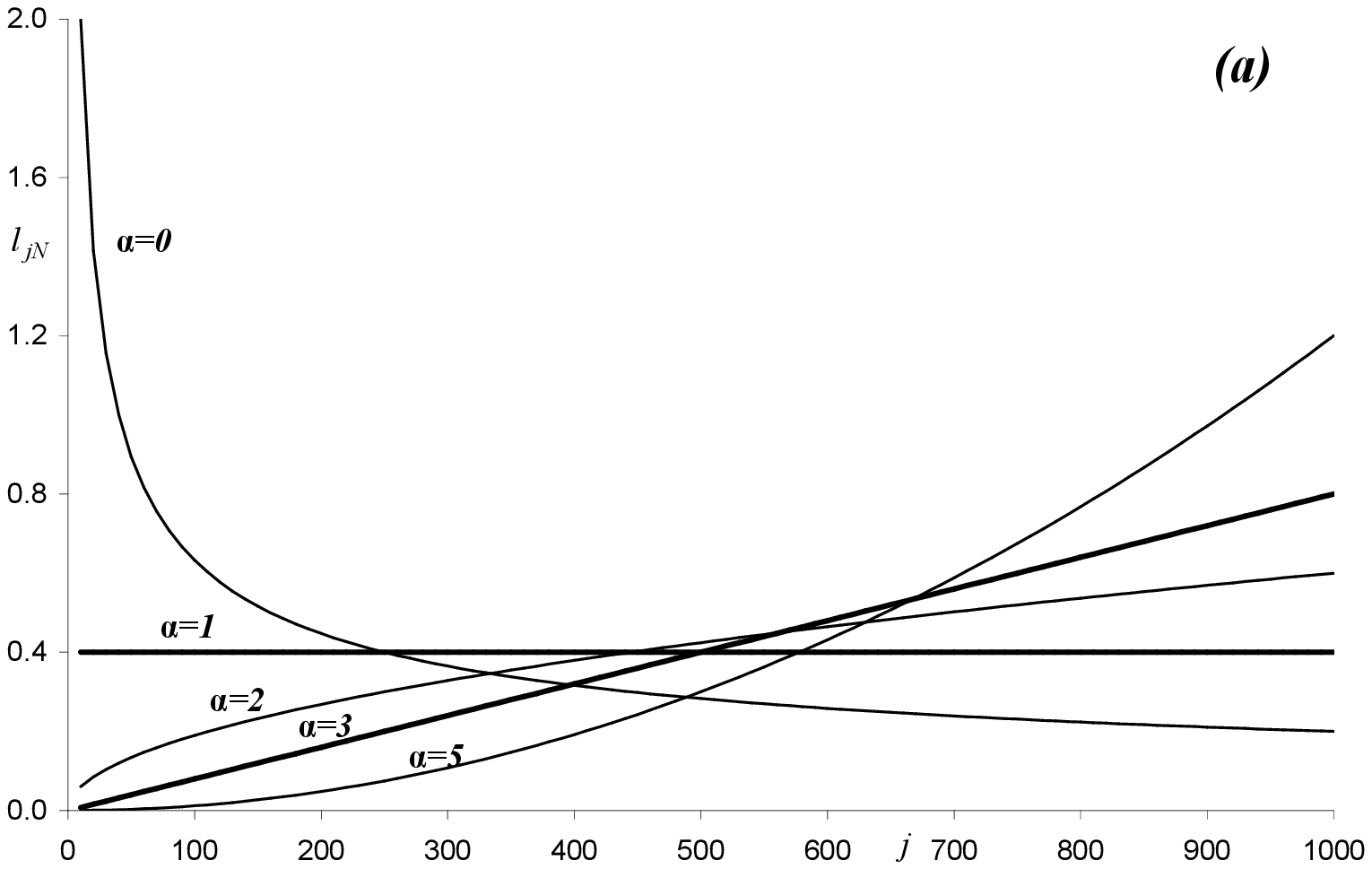}
\includegraphics[width=\columnwidth,clip]{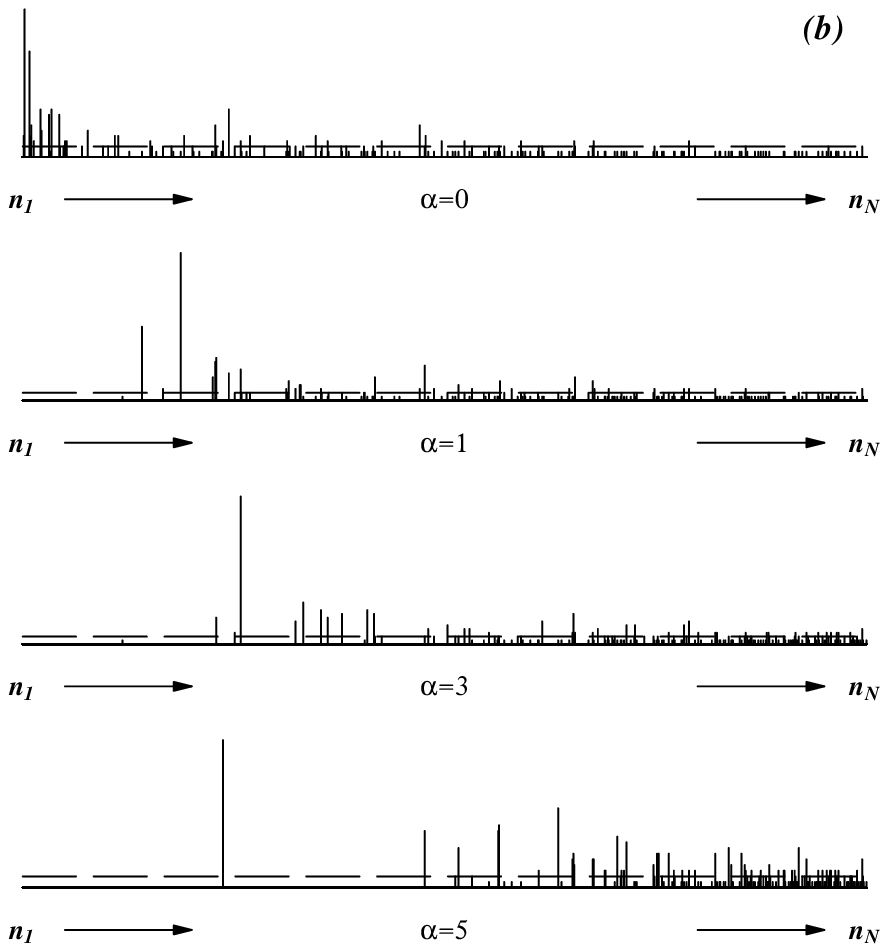}
\caption{\em  Snapshots of the non-stationary statistics of
preferentially attached growing networks all possessing an average
number of links per node $\langle k\rangle=0.4$ when at network
size $N=1000$. (a) The average connectivity distribution $l_{jN}$
as a function of node age $j$ for various values of the
acceleration parameter $\alpha$. Connectivity is monotonically
decreasing for $0\leq\alpha<1$, flat for $\alpha=1$, and
monotonically increasing for $\alpha>1$. (b) Example simulated
networks for various acceleration parameters $\alpha$ with nodes
listed sequentially left to right from $n_1$ to $n_N$ with the
connectivity of each node indicated by vertical lines. The
horizontal dashed line indicates an average connectivity of
two.}\label{f_ljn_dist}
\end{figure}

For convenience, we note that the choice $\beta=0$ sets the
connectivity versus age distribution to be
\begin{equation}     \label{eq_links_l_jn}
  l_{jN} = p  N^{\frac{\alpha+1}{2}} j^{\frac{\alpha-1}{2}},
\end{equation}
for arbitrary $\alpha$. The choice $\alpha=0$ duplicates existing
results found for growing networks which add a constant number of
links with each new node subject to preferential attachment
\cite{Albert_02_47}, while additionally, choosing $p=1$ duplicates
the deterministic results of Ref. \cite{Sen_0310513} in the
regimes $0\leq\alpha<1$ considered in that paper. This purely
preferentially attached distribution is shown in Fig.
\ref{f_ljn_dist} along with example simulation runs at different
acceleration parameters $\alpha$. (The floating free end of the
first link to the first connected node is randomly attached to any
of the existing nodes as they all possess zero links reducing
preferential attachment to random attachment.) Part (a) of this
figure shows the average connectivity $l_{jN}$ of node $n_j$ which
is monotonically decreasing with respect to node age for
$0\leq\alpha<1$ and monotonically increasing when $\alpha>1$.
These different trends depend on whether the accumulative effects
of preferential attachment outweighs the accelerating number of
added links or not. The setting $\alpha=1$ ensures the average
number of links per node is independent of node number as
preferential attachment exactly cancels the bias in initial link
number towards later nodes.  Part (b) of this figure shows the
actual connectivity of node $n_j$ (rather than the average
connectivity) in example simulated networks, making it evident
that actual node connectivity is generally a monotonically
decreasing function of node age.  Why this is so is discussed
later.

The $l_{jN}$ distribution contains information about both node
connectivity and node age and so approximates network statistics
(simulated or observed) when all of this information is available.
However, it is usually the case that node age information is
unavailable necessitating calculation of connectivity
distributions that are not conditioned on node age. This
effectively requires binning together all nodes irrespective of
their age to obtain a final link distribution. Following Ref.
\cite{Gagen_0312021}, the usual continuum approach
\cite{Barabasi_99_17,Barabasi_99_50,Dorogovtsev_01_056125} must be
modified when non-monotonically decreasing connectivity statistics
are present.  In particular, the final connectivity distribution
is obtained via
\begin{eqnarray}         \label{eq_final_link_dist}
  C(k,N) &=& \frac{1}{N} \int_0^N dj \; \delta(k-l_{jN})\nonumber \\
  & = & \pm \frac{1}{N} \left( \frac{\partial l_{jN}}{\partial j}
  \right)^{-1}  \mbox{at }[j=j(k,N)],
\end{eqnarray}
where $j(k,N)$ is the solution of the equation $k=l_{jN}$. The top
line is used when all nodes possess the same average link number
while the second line is applicable with the plus (negative) sign
when the average numbers of links per node is monotonically
increasing (decreasing) with node number. Non-monotonic cases
require alternate approaches.

We now use Eq. \ref{eq_links_l_jn} to calculate the age
independent final link distribution $C(k,N)$ relevant when age
information is unavailable. In the case $\alpha=1$ we have $j\neq
j(k)$, meaning the connectivity $k$ is independent of the node age
$j$, so the delta-function of Eq. \ref{eq_final_link_dist}
immediately integrates to give $\int_0^N dj \;
\delta(k-pN)/N=\delta(k-pN)$. For other cases, we have the
constraint
\begin{equation}
   j = p^{\frac{2}{1-\alpha}} N^{\frac{1+\alpha}{1-\alpha}}
        k^{\frac{2}{\alpha-1}}.
\end{equation}
The age constraint $1\leq j\leq N$ translates into the
connectivity constraints $k\in[pN^\alpha,pN^{(\alpha+1)/2}]$ for
$0\leq\alpha<1$, and $k\in[pN^{(\alpha+1)/2},pN^\alpha]$ for
$\alpha>1$. Via Eq. \ref{eq_final_link_dist}, the final continuous
connection distribution under preferential attachment is then
\begin{equation}    \label{eq_CKN}
  C(k,N) = \left\{
    \begin{array}{ll}
      \frac{2}{1-\alpha} \left(pN^\alpha\right)^{\frac{2}{1-\alpha}}\frac{1}{k^{\frac{3-\alpha}{1-\alpha}}}, &  0\leq\alpha<1, \\
       &  \\
       \delta(k-pN), &   \alpha=1, \\
       &  \\
       \frac{2}{\alpha-1} \frac{k^{\frac{3-\alpha}{\alpha-1}}}{\left(pN^\alpha\right)^{\frac{2}{\alpha-1}}}, &  \alpha>1. \\
    \end{array}
  \right.
\end{equation}
Each of these separate distributions is normalized over
$k\in[k_0,k_{\infty}]$ via $\int_{k_0}^{k_{\infty}} C(k,N)=1$
where $k\in[pN^\alpha,\infty)$ for $0\leq\alpha<1$, $k\in[0,N]$
for $\alpha=1$, and $k\in[0,pN^\alpha]$ for $\alpha>1$. The
average connectivity per node $\langle
k\rangle=\int_{k_0}^{k_{\infty}} k C(k,N)=C/N$ as required over
these same ranges. Again, the choice $p=1$ and $0\leq\alpha<1$
duplicates the deterministic results of Ref. \cite{Sen_0310513}.

\begin{figure}[htbp]
\centering
\includegraphics[width=\columnwidth,clip]{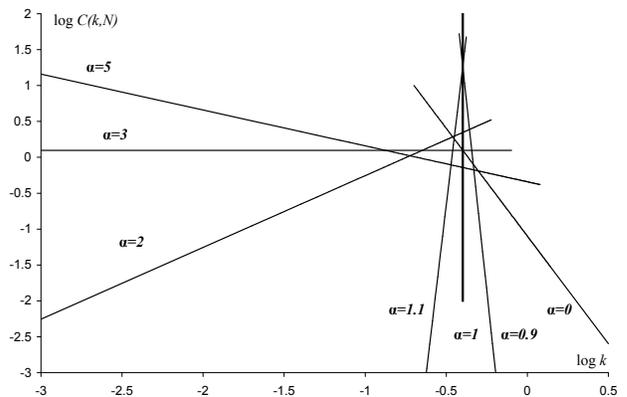}
\caption{\em The age-independent connectivity distribution
$C(k,N)$ as a function of acceleration parameter $\alpha$. The log
plot means that straight lines indicate power-law distributions.
The connectivity distributions are monotonically decreasing for
$0\leq\alpha<1$ over the range $k\in[pN^\alpha,\infty)$, a delta
function for $\alpha=1$, and over the range $k\in[0,pN^\alpha]$,
are monotonically increasing for $1<\alpha<3$, flat for $\alpha=3$
and monotonically decreasing for $\alpha>3$. Networks in this
figure possess an average connectivity per node of $\langle
k\rangle=0.4$ at a network size of $N=10,000$.
}\label{f_ckn_alpha}
\end{figure}

The resulting age-independent connectivity distributions $C(k,N)$
are shown in Fig. \ref{f_ckn_alpha} and are monotonically
decreasing for acceleration parameters $0\leq\alpha<1$,
monotonically increasing for $1<\alpha<3$, uniformly flat for
$\alpha=3$, and monotonically decreasing for $\alpha>3$. In all
cases except $\alpha=1$, the age independent distribution $C(k,N)$
is a power law proportional to $k^{-\gamma}$ with
$\gamma=(\alpha-3)/(\alpha-1)$. The exponent here can take the
values $\gamma\in[3,\infty)$ for $\alpha\in[0,1)$ and
$\gamma\in(-\infty,1)$ for $\alpha\in(1,\infty)$, so no value of
the acceleration parameter permits exponents in the range
$\gamma\in[1,3]$.

The continuous connectivity distributions $C(k,N)$ are typically
defined over very limited ranges $k\in[k_0,k_\infty]$ so it can
often be the case that every node of the network is predicted to
have fractional connectivity less than unity.  Of course, it is
impossible to observe nodes with fractional connectivities---every
node either has zero connections or one connection, or two
connections, and so on. When most connected nodes have fractional
connectivity, it is necessary to convert the unobservable
continuous distribution $C(k,N)$ to an observable discrete
distribution, which we denote $C_k$, giving the proportion of
network nodes possessing an integral number of $k$ links for
$k\in[0,N]$. For networks with connectivity distributions which
are long tailed, the continuous distribution $C(k,N)$ closely
approximates the discrete distribution $C_k$ and this step is
often not required. In Appendix \ref{app_exp_cutoff}, we detail
how to calculate the observed integral connectivity distribution
$C_k$ generated from a continuous power-law connectivity
distribution $C(k,N)$ defined over a finite range
$k\in[k_0,k_\infty]$. It is generally the case that finite ranges
impose exponential cutoffs on scale-free continuous distributions
$C(k,N)$ to give discrete distribution $C_k$ possessing
exponential statistics. A number of alternative mechanisms can
impose an exponential cutoff on a power-law connectivity
distribution
\cite{Dorogovtsev_00_1842,AutoMoreira_02_268703,Mossa_02_138701},
and we now add hyper-acceleration to this list.

\begin{figure}[htbp]
\centering
\includegraphics[width=\columnwidth,clip]{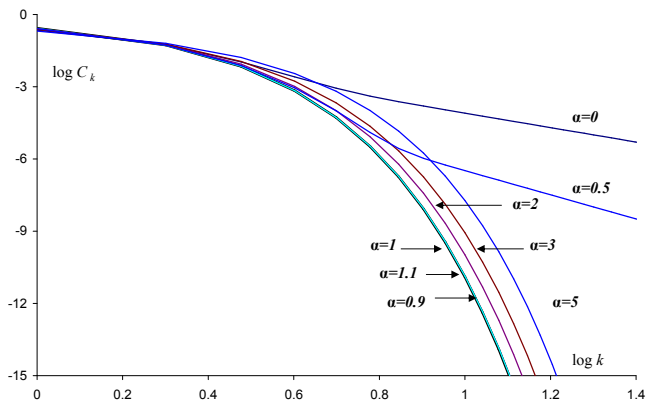}
\caption{\em The age-independent discrete connectivity
distribution $C_k$ as a function of acceleration parameter
$\alpha$ under preferential attachment. It is evident that while
the unobservable continuous distributions $C(k,N)$ appear to be
distinguishable, the observable discrete distributions are
virtually indistinguishable at low connectivity numbers.
}\label{f_ck_discrete_pref}
\end{figure}

Appendix \ref{app_exp_cutoff} specifies a numerical algorithm for
calculating the observable discrete distribution $C_k$ defined by
the unobservable continuous distribution $C(k,N)$. The calculated
discrete distributions equivalent to the continuous distributions
of Fig. \ref{f_ckn_alpha} appear in Fig. \ref{f_ck_discrete_pref}
demonstrating that the generated discrete distributions are
virtually indistinguishable over wide ranges of the acceleration
parameter in marked contrast to the very different shapes of the
continuous distributions $C(k,N)$ at different acceleration
parameters.  This is due to the exponential cutoffs imposed when
the continuous distributions are defined only over finite ranges.
All of the calculations involved in the numerical algorithm of
Appendix \ref{app_exp_cutoff} can in principle be done by hand as
we now illustrate using distributions $C(k,N)$ where the boundary
point $b$ lies within the range of support $k_0<b<k_\infty$ for
low acceleration regimes $0\leq\alpha<1$, but outside the range of
support $k_0<k_\infty<b$ for high acceleration regimes $\alpha\geq
1$. Using Appendix \ref{app_exp_cutoff}, this choice gives
\begin{equation}
  C_k^{+} = \left\{
    \begin{array}{ll}
      \left[\frac{pN^\alpha}{k-1/2}\right]^{\frac{2}{1-\alpha}}-
       \left[\frac{pN^\alpha}{k+1/2}\right]^{\frac{2}{1-\alpha}}, & 0<\alpha<1, k\in[2,\infty), \\
       &  \\
       0, & \alpha=1, k\in[0,N], \\
       &  \\
       0, & \alpha>1. \\
    \end{array}
  \right.
\end{equation}
All of these distributions then have nonzero support to the left
of the boundary $b$ and so can be partitioned to give $NC^-$ nodes
labelled $0\leq j\leq (NC^{-}-1)$ each with average connectivity
$k_j=p_jN$ equal to
\begin{equation}
  k_j = \left\{
    \begin{array}{ll}
      pN^{\alpha} \left(1-\frac{j}{N}\right)^{\frac{\alpha-1}{2}}, & 0<\alpha<1  \\
       &  \\
      pN, & \alpha=1, \\
       &  \\
      pN^{\frac{\alpha}{2}} j^{\frac{\alpha-1}{2}}, & \alpha>1. \\
    \end{array}
  \right.
\end{equation}
These values then feed directly into Eq. \ref{eq_total_C_k} to
give the total discrete distribution $C_k$.  To further illustrate
this last stage of the calculation, consider the case $\alpha=1$
where we have $k_j=p_jN=pN$ and $p_j=p$ for all nodes $0\leq j\leq
N-1\approx N$, and noting the absence of any contribution from the
right of $b$ so $C^+=0$, we have
\begin{equation}
   C_k = C_k^- = {N \choose k} p^k (1-p)^{N-k}, \;\;\;\;\; \alpha=1,
\end{equation}
for $k\in[0,N]$.

As mentioned above, Fig. \ref{f_ljn_dist} makes it apparent that
the actual connectivity of connected nodes generally decreases
with node age even though average node connectivity can be
monotonically increasing or decreasing. The reason is that the
$l_{jN}$ curves average connectivity over all nodes rather than
only over connected nodes. (Recall that $\beta=0$ means that nodes
that are unconnected on entry to the network forever remain
unconnected.) The probability that node $n_j$ is connected is
straightforwardly $P_c(j)=1-P(0,j)$ using Eq. \ref{eq_P_j_k_dist}
or
\begin{equation}
    P_c(j) = 1 - \left( 1- pj^{\alpha-1}\right)^j \;\approx\;  pj^\alpha.
\end{equation}
Hence, the average link number per node at node $n_j$ (Eq.
\ref{eq_links_l_jn}) equates to the product of the average number
of links per connected node at node $n_j$, denoted $l_c(j,N)$, and
the probability that node $n_j$ is connected, or $P_c(j)$. By
definition then, we have
\begin{equation}
  l_{jN}= l_c(j,N) P_c(j),
\end{equation}
giving
\begin{equation}     \label{eq_actual_connectivity}
  l_c(j,N) = \frac{p N^{(\alpha+1)/2} j^{(\alpha-1)/2}}{1 - \left( 1- pj^{\alpha-1}\right)^j}
         \;\approx\;  \left( \frac{N}{j}\right)^{\frac{\alpha+1}{2}}.
\end{equation}
Consequently, the average number of connections of connected nodes
$l_c(j,N)$ is always a monotonically decreasing function of node
age $j$. (An equivalent derivation appeared in Ref.
\cite{Gagen_0312021}.)

\begin{figure}[htbp]
\centering
\includegraphics[width=\columnwidth,clip]{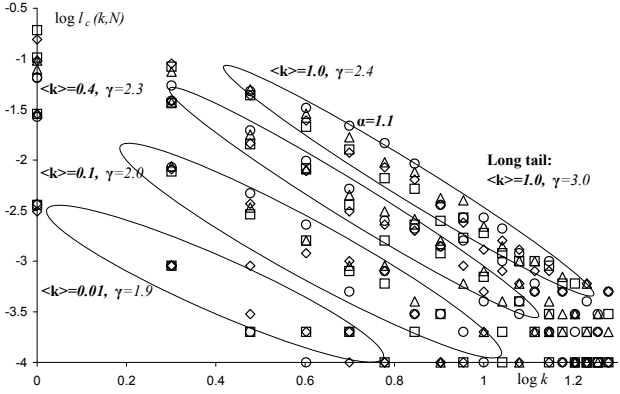}
\caption{\em  The proportional connectivity distribution
$l_c(k,N)$ for connected nodes under preferential attachment for
various values of the average network connectivity $\langle
k\rangle$ at a snapshot network size of $N=10,000$. The curves
naturally group together according to average network
connectivity, and are characterized by a power law $k^{-\gamma}$
with slope $\gamma$ as shown.  Further network growth would
eventually convert these scale-free distributions to exponential
distributions. For average connectivity $\langle k\rangle=0.01$,
we show curves for $(\alpha,p)=(0,0.005),(1,10^{-6}),
(3,2\times10^{-14}), (5,3\times10^{-22})$; for $\langle
k\rangle=0.1$, $(\alpha,p)=(0,0.05), (1,10^{-5}),
(3,2\times10^{-13}), (5,3\times10^{-21})$; for $\langle
k\rangle=0.4$, $(\alpha,p)=(0,0.2), (1,4\times10^{-5}),
(3,8\times10^{-13}),(5,10^{-20})$; for $\langle k\rangle=1.0$,
$(\alpha,p)=(0,0.5),(1,10^{-4}),(3,2\times10^{-12}),
(5,3\times10^{-20})$. }\label{f_k_constant}
\end{figure}

It is difficult to further analyze the connectivity distribution
for actually connected nodes as the distribution of Eq.
\ref{eq_actual_connectivity} is only incompletely sampled---node
$n_j$ is connected only with probability $1-P(0,j)$.  Fig.
\ref{f_k_constant} shows age independent connectivity statistics
for simulated networks at a snapshot size of $N=10,000$ nodes and
with different acceleration and probability parameters
$(\alpha,p)$ chosen to generate networks over a range of average
connectivities $0.01\leq\langle k\rangle\leq 1.0$. To do this, we
choose different acceleration parameters in the set
$\alpha\in\{0,1,3,5\}$ with probability parameters set according
to Eq. \ref{eq_bar_p}. The resulting slopes of the power law
connectivity distributions $k^{-\gamma}$ averaged over the
corresponding acceleration parameters varies over the range
$1.9\leq\gamma\leq 3.0$. This regime was forbidden prior to taking
account of the exponential cutoffs imposed by the finite ranges.
(The choice $(\alpha,p)=(1,10^{-5})$ generating $\gamma\approx 2$
is roughly equivalent to the accelerating simulations of
prokaryote gene regulatory networks \cite{Gagen_0312021}.)
Additional network growth would see these scale-free distributions
undergoing a full transition to exponential statistics.

Finally, we briefly consider the size of the largest connected
island in the artificial case of purely preferentially attached
networks where initially unconnected nodes remain always
unconnected. In this particular case, preferential attachment
ensures that every connected node belongs to the same
interconnected largest island whose size then equates to the
number of connected nodes. As a node is connected only if it gains
some connections on entry to the network with probability
$1-P(0,k)$ from Eq. \ref{eq_P_j_k_dist}, the size $s$ of the
largest island is
\begin{equation}      \label{eq_reg_density}
  s = \sum_{k=1}^{N} \left[ 1- (1-pk^{\alpha-1})^k \right].
\end{equation}
For small $p$ and small network sizes $N$, this equates to the
number of links $s\approx \sum_{k=1}^{N} pk^{\alpha}\approx L$ as
each connected node likely has only one link in sparsely connected
networks.  In this regime, the largest island grows at an
acceleration rate determined by the acceleration parameter
$\alpha$. However, outside this regime, the growth of the largest
island eventually saturates when almost every new node is
connected (i.e. when $1-(1-pk^{\alpha-1})^k\approx 1$) meaning the
largest island grows proportionally to network size $N$. This
occurs in the limit $pk^{\alpha-1}\rightarrow 1$ or $N$ large
giving $s\approx N$. Should a network absolutely need to maintain
an accelerated growth in the size of its largest island, as
conjectured for the regulatory gene network of prokaryotes due to
competitive pressures, then the point at which the largest island
makes a transition from accelerated to linear growth will
represent an upper network size limit
\cite{Gagen_0312021,Gagen_0312022}.

\section{Randomly attached accelerating networks}
\label{sect_random}

We now turn to consider accelerating and hyper-accelerating
networks ($\alpha\geq 0$) where new links are randomly distributed
over all existing nodes.  The final generated link distribution
will depend on the balance of two countervailing trends.  The
first sees newer nodes attracting more links due to acceleration
which confers an increasing average number of links on younger
nodes $n_k$, while in contrast, older nodes have a longer time to
acquire links from the newer nodes.

Under random attachment, the rate of growth of the link number for
node $n_j$ when the network has grown to size $k$ is given under
the continuous approximation
\cite{Barabasi_99_17,Barabasi_99_50,Dorogovtsev_01_056125} as
\begin{equation}    \label{eq_continuum_links_rand}
  \frac{\partial l_{jk}}{\partial k} =  \frac{l_{kk}}{k}
       \;=\; pk^{\alpha-1},
\end{equation}
where $l_{kk}=pk^\alpha$. Here, the rate of growth in the
connectivity $l_{jk}$ is determined by the number of new links
added with node $n_k$ ($l_{kk}$) which are equally proportioned
over the $k$ existing nodes. The resulting connectivity statistics
can readily be obtained through integrating this equation with the
initial conditions $l_{jj}\approx pj^\alpha$ at time $j$ to obtain
\begin{equation}     \label{eq_links_l_jn_rand}
  l_{jN} = \left\{
  \begin{array}{ll}
    p \left[ 1 + \ln \frac{N}{j} \right], & \alpha=0, \\
     &  \\
    \frac{p}{\alpha} \left[ N^\alpha + (\alpha-1) j^\alpha \right], & \alpha>0. \\
  \end{array}
   \right.
\end{equation}
These results satisfy Eq. \ref{eq_continuum_links_rand}, while
integration of the link numbers over all node numbers $j$ gives
the required total number of links as in Eq.
\ref{eq_net_parameters}. (As established in Appendix
\ref{app_pref_random_attach}, these random attachment equations
are exactly reproduced by applying a Taylor series expansion of
the general equations Eq. \ref{eq_links_l_jn_beta} about the point
$x=pN^\alpha/\beta\approx 0$ with the retention only of terms
linear in $x$.)

In passing, we present the exact distribution for links under
random attachment.  First, note that node $n_k$ initially receives
an average of $\langle j\rangle=pk^\alpha$ links where $j\in[0,k]$
with each link formed with probability $pk^{\alpha-1}$, and
subsequently receives an average of $\langle
j_k\rangle=pk^{\alpha}/k=pk^{\alpha-1}$ links from itself where
$j_k\in[0,k]$ with each link formed with probability
$pk^{\alpha-1}$. The arrival of node $n_{k+1}$ is accompanied by
an additional $p(k+1)^{\alpha}$ new links of which an average
$\langle j_{k+1}\rangle=p(k+1)^{\alpha}/(k+1)=p(k+1)^{\alpha-1}$
attach to node $n_k$ where $j_{k+1}\in[0,k+1]$ with each link
formed with probability $p(k+1)^{\alpha-2}$, and so on until it
receives an average of $\langle
j_{N}\rangle=pN^{\alpha}/N=pN^{\alpha-1}$ links from node $n_N$
with $j_N\in[0,N]$ with each link formed with probability
$pN^{\alpha-2}$. The joint probability that node $n_k$ obtains
$j,j_k,\dots,j_N$ connections is then
\begin{eqnarray}              \label{eq_pjjkjk1_etc}
   P_k(j,j_k,j_{k+1},\dots,j_N) &=&   \\
   && \hspace{-2cm} {k \choose j} \left[pk^{\alpha-1}\right]^j
   \left[1-pk^{\alpha-1}\right]^{k-j} \times \nonumber \\
   && \hspace{-2cm} \times \prod_{n=k}^N
     {n \choose j_n} \left[pn^{\alpha-2}\right]^{j_n}
     \left[1-pn^{\alpha-2} \right]^{n-j_n}. \nonumber
\end{eqnarray}
The average number of inbound links for node $n_k$ is then
\begin{equation}
   \langle j+j_k+\dots+j_N\rangle
           = pk^{\alpha} + \sum_{n=k}^N pn^{\alpha-1},
\end{equation}
which integrates to give Eq. \ref{eq_links_l_jn_rand} as found by
the continuum approach. Unfortunately, further work with this more
exact multivariate distribution is generally intractable.

\begin{figure}[htbp]
\centering
\includegraphics[width=\columnwidth,clip]{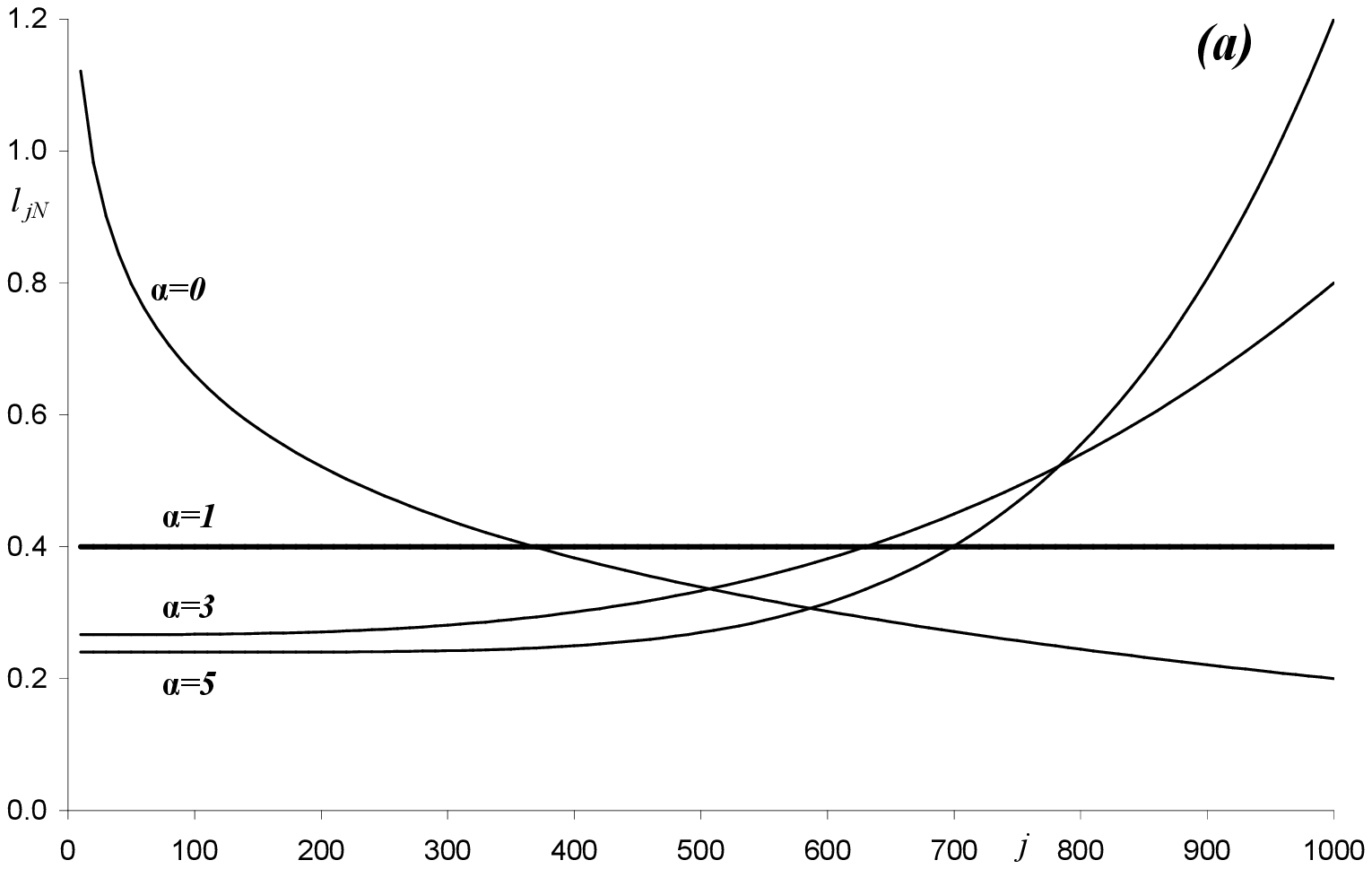}
\includegraphics[width=\columnwidth,clip]{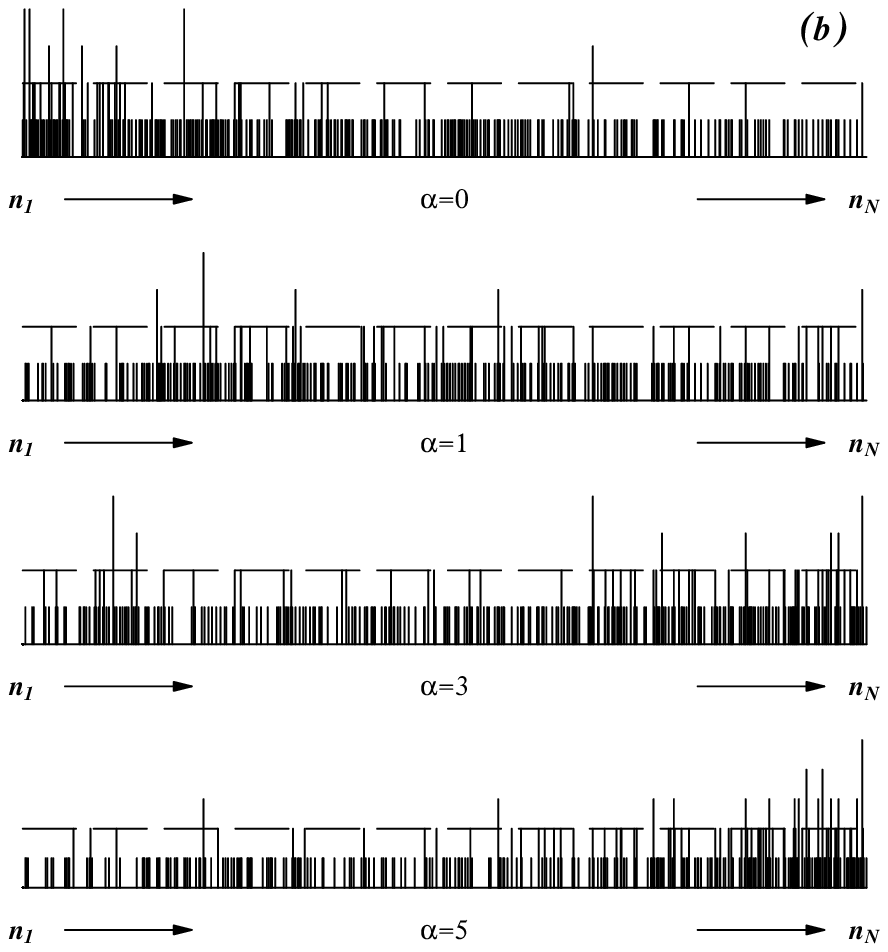}
\caption{\em  (a) The average connectivity distribution $l_{jN}$
under random attachment as a function of node age $j$ for various
values of the acceleration parameter $\alpha$.  Connectivity is
monotonically decreasing for $0\leq\alpha<1$, flat for $\alpha=1$,
and monotonically increasing for $\alpha>1$. (b) Example simulated
networks for various acceleration parameters $\alpha$ with nodes
listed sequentially left to right from $n_1$ to $n_N$ with the
connectivity of each node indicated by vertical lines. The
horizontal dashed line indicates an average connectivity of two.
}\label{f_ljn_dist_rand}
\end{figure}

The resulting connectivity $l_{jN}$ verses node age $j$
distributions are shown in Fig. \ref{f_ljn_dist_rand} along with
example simulation runs at different acceleration parameters
$\alpha$.  This figure makes clear that the connectivity
distribution $l_{jN}$ is monotonically decreasing with $j$ for
$0\leq\alpha< 1$ so younger nodes are more heavily connected than
older nodes on average, and monotonically increasing for
$\alpha>1$ so younger nodes are less heavily connected than older
nodes on average.  It is only on the boundary between these two
regions at $\alpha=1$ that the connectivity distribution is flat
so average node connectivity is independent of node age.

\begin{figure}[htbp]
\centering
\includegraphics[width=\columnwidth,clip]{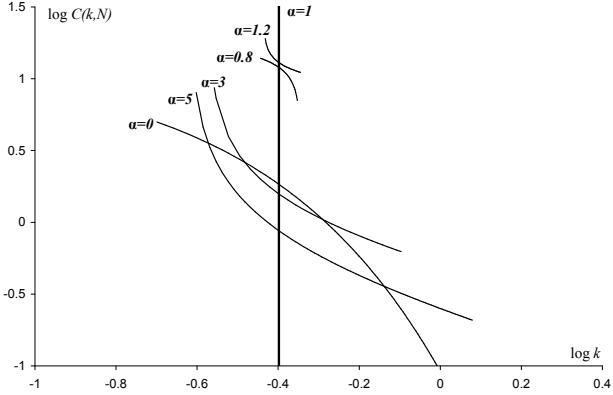}
\caption{\em The age-independent connectivity distribution
$C(k,N)$ as a function of acceleration parameter $\alpha$ for
randomly connected networks. The distributions are monotonically
decreasing for $\alpha\neq 1$ over the respective ranges
$k\in[p,\infty)$ for $\alpha=0$,
$k\in[pN^\alpha,pN^\alpha/\alpha]$ for $0<\alpha<1$, and
$k\in[pN^\alpha/\alpha,pN^\alpha]$ for $\alpha>1$.
}\label{f_ckn_alpha_rand}
\end{figure}

As previously, the $l_{jN}$ distribution contains information
about both node connectivity and node age and so approximates
network statistics (simulated or observed) when all of this
information is available. We now calculate the age independent
final link distribution $C(k,N)$ as in Eq.
\ref{eq_final_link_dist}.  In the case $\alpha=1$ we have $j\neq
j(k)$, meaning the connectivity $k$ is independent of the node age
$j$, so the delta-function of Eq. \ref{eq_final_link_dist} again
integrates to give $\delta(k-pN)$. For other cases, we have
\begin{equation}
   j = \left\{
   \begin{array}{ll}
     N e^{(1-k/p)},& \alpha=0, \\
      &  \\
     \left[\frac{1}{1-\alpha} \left(N^\alpha - \frac{\alpha k}{p} \right) \right]^{1/\alpha},  & 0<\alpha, \alpha\neq 1. \\
   \end{array}
   \right.
\end{equation}
The age constraint $1\leq j\leq N$ translates into the
connectivity constraints $k\in[p,p(1+\ln N)]$ for $\alpha=0$,
$k\in[pN^\alpha,pN^\alpha/\alpha]$ for $0<\alpha<1$, $k=pN$ for
$\alpha=1$, and $k\in[pN^\alpha/\alpha,pN^\alpha]$ for $\alpha>1$.
Via Eq. \ref{eq_final_link_dist}, these relations determine the
predicted age independent final link distribution $C(k,N)$ under
random attachment to be
\begin{eqnarray}    \label{eq_CKN_random}
  C(k,N) &=& \\
  && \hspace{-1cm} \left\{
    \begin{array}{ll}
      \frac{1}{p} e^{(1-k/p)}, & \alpha=0, \\
       &   \\
      \frac{1}{pN(1-\alpha)^{1/\alpha}} \left(N^\alpha - \frac{\alpha k}{p} \right)^{\frac{1-\alpha}{\alpha}},  & 0\leq\alpha<1, \\
       &   \\
       \delta(k-pN),   &  \alpha=1, \\
       &  \\
      \frac{1}{pN(\alpha-1)^{1/\alpha}} \frac{1}{\left(\frac{\alpha k}{p}-N^\alpha \right)^{\frac{\alpha-1}{\alpha}}},  & \alpha>1. \\
    \end{array}
  \right.  \nonumber
\end{eqnarray}
Each of these separate distributions is normalized over
$k\in[k_0,k_{\infty})$ via $\int_{k_0}^{k_{\infty}} C(k,N)=1$
where $k\in[p,\infty)$ for $\alpha=0$,
$k\in[pN^\alpha,pN^\alpha/\alpha]$ for $0<\alpha<1$, $k=pN$ for
$\alpha=1$, and $k\in[pN^\alpha/\alpha,pN^\alpha]$ for $\alpha>1$.
The average connectivity per node $\langle
k\rangle=\int_{k_0}^{k_{\infty}} k C(k,N)=C/N$ as required over
these same ranges. As shown in Fig. \ref{f_ckn_alpha_rand}, the
predicted age independent continuous connectivity distribution
$C(k,N)$ is monotonically decreasing as a function of $k$ for both
$\alpha<1$ and $\alpha>1$, and a delta-function for $\alpha=1$.

\begin{figure}[htbp]
\centering
\includegraphics[width=\columnwidth,clip]{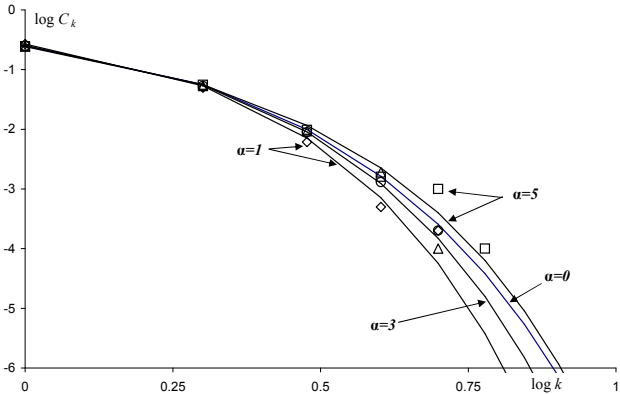}
\caption{\em The age-independent discrete connectivity
distribution $C_k$ as a function of acceleration parameter
$\alpha$ for random attachment. It is evident that while the
unobservable continuous distributions $C(k,N)$ are
distinguishable, the observable discrete distributions are
virtually indistinguishable. We also show the actual points for
different network simulations of $N=10,000$ nodes with average
connectivity $\langle k\rangle=0.4$ with $\alpha=0$ (triangles),
$\alpha=1$ (diamonds), $\alpha=3$ (circles), and $\alpha=5$
(squares).}\label{f_discrete_rand}
\end{figure}

The continuous connectivity distributions $C(k,N)$ are again
typically defined over very limited ranges $k\in[k_0,k_\infty]$
for acceleration parameters greater than zero so many nodes
possess fractional connectivity less than unity. This again
requires that the continuous distribution $C(k,N)$ be converted
into an equivalent discrete distribution $C_k$ following the
methods of Appendix \ref{app_exp_cutoff}.  Fig.
\ref{f_discrete_rand} shows the results of taking full account of
the finite ranges of the continuous distributions $C(k,N)$ which
impose exponential cutoffs on the generated discrete distribution
$C_k$. It is clear that the exponential cutoffs render the
observable discrete distributions for different acceleration
parameters essentially indistinguishable.  This result is borne
out by simulations of networks with $N=10,000$ nodes with average
connectivity $\langle k\rangle=0.4$ for different acceleration
parameters (also shown in Fig. \ref{f_discrete_rand}).

Again, we illustrate the numerical algorithm of Appendix
\ref{app_exp_cutoff} via specific examples for randomly attached
networks.  Consider continuous distributions $C(k,N)$ where the
boundary point $b$ lies within the range of support for the
distribution for all acceleration parameters except $\alpha=1$.
That is, we assume $pN^\alpha<b<pN^\alpha/\alpha$ for
$0\leq\alpha<1$ and $pN^\alpha/\alpha<b<pN^\alpha$ for $\alpha>1$,
while for $\alpha=1$ we assume $pN<b$.  These illustrative choices
then give
\begin{equation}
  C_k^{+} = \left\{
    \begin{array}{ll}
      2 \sinh\left( \frac{1}{2p} \right) e^{(1-k/p)}, & \alpha=0,  \\
       &  \\
  \left[\frac{1-\frac{\alpha}{pN^\alpha}(k-\frac{1}{2})}{1-\alpha}\right]^{\frac{1}{\alpha}}-
  \left[\frac{1-\frac{\alpha}{pN^\alpha}(k+\frac{1}{2})}{1-\alpha}\right]^{\frac{1}{\alpha}}, & 0<\alpha<1,  \\
       &  \\
      0, & \alpha=1, \\
       &  \\
  \left[\frac{\frac{\alpha}{pN^\alpha}(k+\frac{1}{2})-1}{\alpha-1}\right]^{\frac{1}{\alpha}}-
  \left[\frac{\frac{\alpha}{pN^\alpha}(k-\frac{1}{2})-1}{\alpha-1}\right]^{\frac{1}{\alpha}}, & \alpha>1. \\
    \end{array}
  \right.
\end{equation}
In this example, the allowed ranges for the connectivity $k$ are
$k\in[2,\infty)$ for $\alpha=0$, $k\in[2,pN^\alpha/\alpha]$ for
$0<\alpha<1$, and $k\in[2,pN^\alpha]$ for $\alpha>1$.  Our example
distributions have nonzero support to the left of the boundary
point $b$ and so can be partitioned to give $NC^-$ nodes labelled
$0\leq j\leq (NC^{-}-1)$ each with average connectivity $k_j=p_jN$
equal to
\begin{equation}
  k_j = \left\{
    \begin{array}{ll}
      p \left[ 1 - \ln \left(1-\frac{j}{N}\right)\right], & \alpha=0, \\
       &  \\
       \frac{pN^\alpha}{\alpha} \left[ 1 -
  (1-\alpha) \left(1-\frac{j}{N} \right)^\alpha
  \right], & 0<\alpha<1,  \\
       &  \\
      pN, & \alpha=1, \\
       &  \\
      \frac{pN^\alpha}{\alpha} \left[ 1 +
  (\alpha-1) \left(\frac{j}{N} \right)^\alpha
  \right], & \alpha>1. \\
    \end{array}
  \right.
\end{equation}
These values then feed directly into Eq. \ref{eq_total_C_k} to
give the total discrete distribution $C_k$.  To illustrate this
last stage of the calculation, consider the case $\alpha=1$ where
we have $k_j=p_jN=pN$ and $p_j=p$ for all nodes $0\leq j\leq
N-1\approx N$, and noting the absence of any contribution from the
right of $b$ so $C^+=0$, we have
\begin{equation}
   C_k = C_k^- = {N \choose k} p^k (1-p)^{N-k}, \;\;\;\;\; \alpha=1,
\end{equation}
for $k\in[0,N]$.

\begin{figure}[htbp]
\centering
\includegraphics[width=\columnwidth,clip]{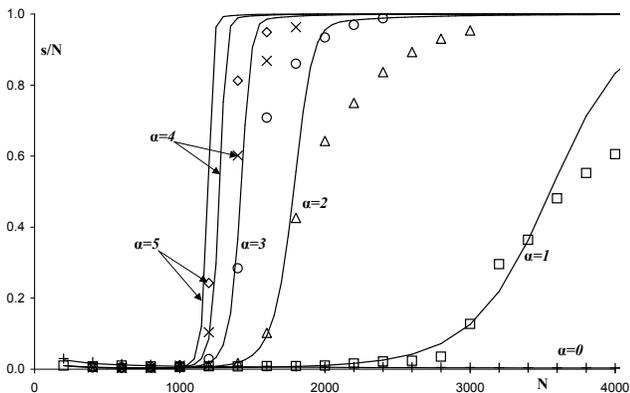}
\caption{\em The simulated and predicted size of the largest
island $s$ expressed as a proportion of growing network size for
various acceleration parameters $\alpha$.  Each network has
average connectivity $\langle k\rangle=0.4$ at a critical network
size $N_c=1000$ nodes, and in every case, we assume that the
average size of smaller islands is $\bar{s}=7.5$.
}\label{f_islands}
\end{figure}

We now turn to establish how the formation of a single giant
connected component depends on the acceleration parameter
$\alpha$.  The number of nodes in the largest island $s$ will be a
function of network size and will grow as new nodes and new
connections are added to the network.  The largest island will
grow by one when the new node $n_k$ forms one link with
probability $P(1,k)$ which attaches to the largest island with
probability $s/k$, and will grow by the average size of all
smaller islands $\bar{s}$ when node $n_k$ forms two links with
probability $P(2,k)$ either of which connects to the largest
island with probability $s/k$ while the remaining link connects to
an external island with probability $(k-s)/k$. Altogether, the
rate of growth of the largest island goes as
\begin{eqnarray}
  \frac{ds}{dk} &=& \frac{s}{k} P(1,k) +
        2 \bar{s} \frac{(k-s)s}{k^2} P(2,k)  \nonumber \\
    &\approx& p k^{\alpha-1}s +
          p^2 \bar{s} k^{2\alpha-2} (k-s) s,
\end{eqnarray}
using Eq. \ref{eq_P_j_k_dist} and noting the restriction $p
k^{\alpha-1}\leq 1$. Numerical simulations (and some analytically
tractable solutions) indicate that the first term here is
negligible compared to the second, and hereinafter we ignore this
first term. For initial conditions, we have $s(n_c)=2$ at node
$n_c$ where the first link likely appears, see Appendix
\ref{app_exp_cutoff}.  Fig. \ref{f_islands} shows the predicted
growth of the largest island compared to simulated networks for
different acceleration parmeters $\alpha$, where all networks
possess an average connectivity per node of $\langle l\rangle=0.4$
at a critical network size of $N=1000$, and we assume an average
size of smaller islands of $\bar{s}=7.5$. There is an evident
close connection between theoretically predicted curves and
observed statistics.

\section{Conclusion}

We have examined the network structural transitions displayed by
accelerating and hyper-accelerating probabilistic networks
motivated by the observation that important biological networks
such as prokaryote gene regulatory networks are both
hyper-accelerating and size limited due to network structural
transitions.  We examined accelerating growing networks of nodes
connected by undirected links which were probabilistically added
with each new node and either preferentially or randomly attached
to existing nodes.  The addition of a probabilistic envelope
allowing the number of new links added with each new node to be an
integer greater than or equal to zero, allowed us to extend
network theory to model sparsely connected networks where the
majority of nodes are entirely unconnected (e.g. prokaryote gene
regulatory networks). Consequently, the probabilistic envelope
also allowed us to lift existing constraints on the modelling of
accelerating networks allowing us to treat decelerating,
accelerating, and hyper-accelerating networks.  These latter two
classes of networks were shown to be subject to transitions in
which either a single giant connected component forms or the
network condenses into a fully connected state with exponential
statistics.  We were able to roughly locate these transitions as a
function of the acceleration and probability parameters and the
network size.  Mean field approximations were compared to network
simulations over a wide range of parameters and shown to be
consistent.

The present paper uses mean field network theory to model rapidly
accelerating networks consisting of many unconnected nodes
allowing the examination of the statistical transitions generated
under accelerated growth.  As such, these extended models will be
useful in treating models of, for instance, accelerating
biological regulatory networks mainly consisting of unconnected
nodes and displaying network transitions which limit size and
complexity.  Such models are required to explain the observed
evolutionary record of prokaryote gene regulatory networks.

\appendix
\section{Relative weights of preferential and random attachment}
\label{app_pref_random_attach}

In analyzing Eq. \ref{eq_links_l_jn_beta}, it is first necessary
to determine appropriate limits on $\beta$ giving access to
regimes where preferential attachment dominates random attachment
for accelerating networks.

Appropriate limits on $\beta$ can be obtained by performing a
Taylor expansion of Eq. \ref{eq_links_l_jn_beta} about the point
$x=pN^\alpha/\beta\approx 0$ while retaining only terms linear in
$x$.  Using $\alpha=0$ in Eq. \ref{eq_links_l_jn_beta}, we have
\begin{equation}
   l_{jN} = \frac{p}{x} \left[ (1+x) \left(\frac{N}{j} \right)^x -1
   \right].
\end{equation}
Noting $d/dx [(1+x)a^x]=a^x[1+(1+x)\ln a]$, the leading terms of a
Taylor expansion immediately reproduce the $\alpha=0$ result for
the random attachment model of Eq. \ref{eq_links_l_jn_rand}.
Similarly, using $\alpha>0$ in Eq. \ref{eq_links_l_jn_beta} gives
\begin{equation}
  l_{jN} = \beta \left\{
       \left[1+x \left(\frac{j}{N} \right)^\alpha \right]
       \left[
         \frac{\frac{2}{\alpha+1}x^\alpha+1}{\frac{2}{\alpha+1}x \left(\frac{j}{N} \right)^\alpha+1}
       \right]^{\frac{\alpha+1}{2\alpha}}
       - 1 \right\}.
\end{equation}
A straightforward differentiation then gives a Taylor expansion
whose leading terms exactly equal the $\alpha>0$ result for the
random attachment model of Eq. \ref{eq_links_l_jn_rand}. Hence,
random attachment entirely dominates when $pN^\alpha/\beta\ll 1$
or equivalently, when
\begin{equation}
  \beta \gg \beta_{\rm random} = pN^\alpha\;=\; \frac{1}{2}
  (1+\alpha)\langle k\rangle.
\end{equation}
For values of $\beta<\beta_{\rm random}$, preferential attachment
will influence the final distribution.  For sufficiently small
$\beta$, preferential attachment will dominate (rather than merely
contribute) and this latter boundary can be located as follows.
The first connected node, denoted $n_c$ (i.e. the $c^{\rm th}$
node), likely appears when the cumulative average number of added
initial links sums roughly to unity,
\begin{equation}
  \sum_{k=1}^{c} pk^\alpha \approx
    \frac{pc^{\alpha+1}}{\alpha+1} \approx 1.
\end{equation}
Now the floating end of the new link attached to node $n_c$ can be
either randomly attached to one of the nodes $n_1$ to $n_c$ with
probability proportional to $c\beta$, or preferentially attached
to node $n_c$ with probability proportional to $1+\beta\approx 1$
in the preferential attachment regime. For preferential attachment
to dominate, we require $c\beta\ll 1$, or equivalently,
\begin{equation}
  \beta \ll \beta_{\rm pref} = \frac{1}{c} =
  \left[\frac{p}{\alpha+1}\right]^{\frac{1}{\alpha+1}}
     \; = \; \left[ \frac{\langle k\rangle}{2 N_c^\alpha}\right]^{\frac{1}{\alpha+1}}.
\end{equation}
For the typical hyper-accelerating networks considered here, this
last constraint ensures that preferential attachment dominates
only when $\beta$ is so small as to be essentially zero, and in
this paper, we analyze preferential attachment using the setting
$\beta=0$.

\begin{figure}[htbp]
\centering
\includegraphics[width=\columnwidth,clip]{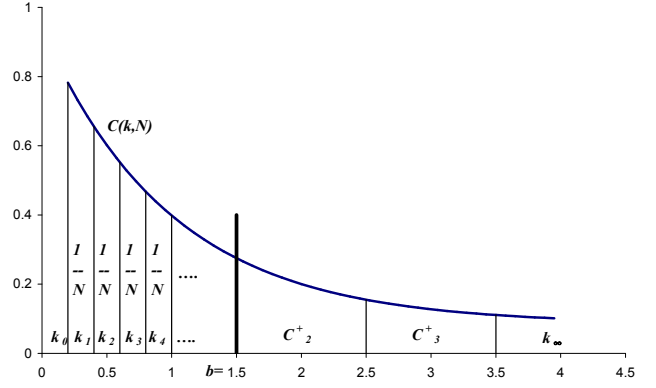}
\caption{\em A continuous distribution $C(k,N)$ defined over the
finite range $[k_0,k_\infty]$ generates an observable discrete
distribution $C_k$ for $k=0,1,2,\dots$.  The $C(k,N)$ distribution
lying to the left of the boundary $b=3/2$ represents network nodes
with fractional average connectivity distributions, and is
partitioned into regions each of area $1/N$ with boundaries
$[k_j,k_{j+1})$ for $j\geq 0$.  The $C(k,N)$ distribution lying to
the right of the boundary $b=3/2$ represents network nodes with
integral average connectivity, and is partitioned into regions as
shown. }\label{f_cont_discrete}
\end{figure}

\section{Exponential cutoffs in hyper-accelerating networks}
\label{app_exp_cutoff}

An arbitrary continuous network connectivity distribution $C(k,N)$
defined over a finite range $[k_0,k_\infty]$ is essentially
unobservable, and in actuality, generates an equivalent discrete
connectivity distribution $C_k$ with $k\in[0,N]$ which can be
compared to observed statistics. Here, we show how to calculate
the observable discrete distribution from the theoretically
predicted but unobservable continuous distribution.

The first step is to partition the continuous distribution
$C(k,N)$ into two parts at an arbitrarily chosen boundary point
$b$ of order unity.  We choose $b=3/2$ as shown in Fig.
\ref{f_cont_discrete}. Then, that proportion of the continuous
distribution $C(k,N)$ lying to the right of $b$ can be considered
to contribute to the long tail of the observed discrete
distribution $C_k$ in the normal way. That is, the proportion of
the continuous distribution in the region $[k-1/2,k+1/2)$ for
integral $k$ equal to
\begin{equation}   \label{eq_L_plus}
   C_k^+ = \int_{k-1/2}^{k+1/2} C(k,N) \; dk,
\end{equation}
can be entirely assigned to the discrete distribution bin $C_k$.
The total proportion of the continuous distribution $C(k,N)$ so
assigned is
\begin{equation}
   C^+  =  \sum_{k=b+1/2}^{k_{\infty}} C_k^+ \;=\;
            \int_{b}^{k_{\infty}} C(k,N) \; dk,
\end{equation}
and this proportion of the distribution describes the connectivity
of a total of $NC^+$ nodes.  When this proportion is close to
unity as usually applies for long tailed distributions, nothing
further need be done and the distribution bins $C_k^+=C_k$ then
equate to the predicted discrete observable distribution.

However, when $C^+$ is significantly less than unity as applies
for hyper-accelerating networks, it is necessary to assign the
remaining proportion of the continuous distribution lying to the
left of the chosen value $b$ to the discrete distribution bins
$C_k$. The proportion of the continuous distribution remaining to
be assigned is
\begin{equation}       \label{eq_L_minus_limit}
   C^-  =  \int_{k_0}^{b} C(k,N)\; dk \;=\; 1 - C^+,
\end{equation}
with this proportion of the distribution describing the
connectivity of a total of $NC^-$ nodes. The best way to
understand how this remaining assignment is done is through a
successive partitioning of the usual normalization constraint
\begin{eqnarray}
  1 &=& \int_{k_0}^{k_{\infty}} C(k,N) \; dk \nonumber \\
  & = & \int_{k_0}^{b} C(k,N) \; dk + \int_{b}^{k_{\infty}} C(k,N) \; dk  \nonumber \\
  & = & \sum_{j=0}^{(NC^{-}-1)} \int_{k_j}^{k_{j+1}} C(k,N) \; dk +
         C^+.
\end{eqnarray}
Here, the continuous distribution to the left of the point $k=b$
is partitioned into $NC^-$ regions with respective boundaries
$[k_j,k_{j+1})$ for $k_j\in[k_0,b]$ and $k_j<k_{j+1}$ while as
noted above, that to the right of $k=b$ is partitioned into
potentially an infinite number of regions with boundaries
$[k-1/2,k+1/2)$ for integral $k>b$. The second contribution to the
discrete bin $C_k$ comes from the $NC^-$ regions of the continuous
distribution lying to the left of $b$. We would like each of these
$NC^-$ regions to correspond to a single node of the network with
non-integral average connectivity, and achieve this by choosing
the regional boundaries $k_j$ such that
\begin{equation}     \label{eq_k_boundary}
   \int_{k_0}^{k_j} C(k,N) \; dk = \frac{j}{N}, \;\;\;\;\;
   j\in[0,NC^{-}-1].
\end{equation}
Here, each region defines an area of exactly $1/N$ so $j$ regions
have a combined area of $j/N$, see Fig. \ref{f_cont_discrete}. The
average connectivity of the node corresponding to the $j^{\rm th}$
region is approximately
\begin{equation}       \label{eq_kj_average}
   \langle k_j\rangle= k_j,
\end{equation}
and is typically non-integral. (Other alternatives such as
$\langle k_j\rangle=(k_j+k_{j-1})/2$ might be chosen but the
difference is negligible for large $N$.) As any given node can, in
practise, only possess an integral number of connections, these
non-integral average connectivities require that the link numbers
for these nodes be assigned probabilistically over the range
$k\in[0,N]$ so as to give the required average. This is achieved
by assigning an integral number $l_j$ of links to this region with
$l_j\in[0,N]$ according to the probability distribution
\begin{equation}
   P_r(l_j) = {N \choose l_j} p_j^{l_j} (1-p_j)^{N-l_j},
\end{equation}
with probabilities $p_j$ chosen to satisfy
\begin{equation}       \label{eq_pj}
   p_j=\frac{\langle k_j\rangle}{N},
\end{equation}
to ensure that this distribution has average $\langle
l_j\rangle=p_jN=\langle k_j\rangle$. We note that later results
will not be overly dependent on whether the $P_r$ distribution
chosen here is binomial in form, or some other reasonable
distribution. Summing over all the $NC^-$ regions to the left of
$b$, the proportion of nodes possessing an integral number of
$l_j=k$ links is then
\begin{equation}    \label{eq_L_minus}
   C_k^- = \frac{1}{N} \sum_{j=0}^{(NC^{-}-1)} {N \choose k} p_j^{k}
   (1-p_j)^{N-k},
\end{equation}
for $k\in[0,N]$ and zero otherwise. This is normalized according
to $\sum_{k=0}^{N}C_k^-=C^-$. Consequently, the total proportion
of the network possessing a discrete number of $k$ links is
expected to be
\begin{equation}
   C_k = C_k^- + C_k^+.
\end{equation}
As required, this distribution is normalized and has average
$\langle k \rangle=L/N$. This gives the discrete probability that
a node has $k$ links for $k\in[0,\infty)$.  It is also common to
calculate the related probability that connected nodes have $k$
links and this is given by $C'_k=C_k/(1-C_0)$ for $k\geq 1$.

In summary, for the general case of arbitrary continuous
distributions $C(k,N)$ defined over the range
$k\in[k_0,k_\infty]$, the predicted discrete distribution is
\begin{eqnarray}    \label{eq_total_C_k}
 C_k &=&  C_k^- + C_k^+    \nonumber \\
 &=& \frac{1}{N} \sum_{j=0}^{NC^{-}-1} {N \choose k} p_j^k
             (1-p_j)^{N-k}  \nonumber \\
 && \hspace{1cm} + \int_{k-1/2}^{k+1/2} C(k,N) \; dk,
\end{eqnarray}
where $C^-=\int_{k_0}^{b} C(k,N) \; dk$, $p_j=k_j/N$, $k_j$ is
chosen to satisfy $j/N=\int_{k_0}^{k_j} C(k,N) \; dk$, and the
partition point $b$ is arbitrarily chosen to be close to unity.

\bibliographystyle{unsrt}

\end{document}